\documentclass[a4paper,aps,prd,10pt,preprintnumbers,showpacs,twocolumn,superscriptaddress,nofootinbib,amsmath,amssymb]{revtex4-1}\def\PRDstyle#1{#1}\def\JCAPstyle#1{}\def\Abstract#1{\begin{abstract}#1\end{abstract}}
\usepackage{graphicx}
\usepackage{cmap}
\usepackage[utf8]{inputenc}
\usepackage[T1]{fontenc}

\def\imo{i}

\def\Order#1{{\cal O}\left(#1\right)}

\newcommand{\ie}{{i.e.,}~}
\newcommand{\eg}{{e.g.,}~}

\begin{document}
\title{General black-hole metric mimicking Schwarzschild spacetime}

\PRDstyle{
\author{R. A. Konoplya} \email{roman.konoplya@gmail.com}
\affiliation{Institute of Physics and Research Centre of Theoretical Physics and Astrophysics, Faculty of Philosophy and Science, Silesian University in Opava, CZ-746 01 Opava, Czech Republic}
\author{A. Zhidenko} \email{olexandr.zhydenko@ufabc.edu.br}
\affiliation{Institute of Physics and Research Centre of Theoretical Physics and Astrophysics, Faculty of Philosophy and Science, Silesian University in Opava, CZ-746 01 Opava, Czech Republic}
\affiliation{Centro de Matemática, Computação e Cognição (CMCC), Universidade Federal do ABC (UFABC),\\ Rua Abolição, CEP: 09210-180, Santo André, SP, Brazil}
}

\JCAPstyle{
\author[\dagger]{R. A. Konoplya,}\emailAdd{roman.konoplya@gmail.com}
\author[\dagger\ddagger]{A. Zhidenko}\emailAdd{olexandr.zhydenko@ufabc.edu.br}
\affiliation[\dagger]{Institute of Physics and Research Centre of Theoretical Physics and Astrophysics,\\ Faculty of Philosophy and Science, Silesian University in Opava,\\ CZ-746 01 Opava, Czech Republic}
\affiliation[\ddagger]{Centro de Matemática, Computação e Cognição (CMCC),\\ Universidade Federal do ABC (UFABC),\\ Rua Abolição, CEP: 09210-180, Santo André, SP, Brazil}
\arxivnumber{2303.03130}
}

\Abstract{
Using the general parametrization of spherically symmetric and asymptotically flat black holes in arbitrary metric theories of gravity and implying that: a) the post-Newtonian constraints are taken into account and b) basic astrophysically relevant characteristics (such as, dominant quasinormal modes, frequency at the innermost stable circular orbit, binding energy, radius of the shadow etc.) are indistinguishable from their Schwarzschild values, we propose a simple metric which depends on three independent parameters (coefficients of the parametrization). Variation of these three parameters can, nevertheless, lead to the two distinctive features. The first is the black-hole temperature, and consequently the Hawking radiation, which can differ a lot from its Schwarzschild limit. The second is the outburst of overtones which become extremely sensitive to small changes of the parameters.
}

\maketitle

\section{Introduction}
A black hole in the Einstein theory of gravity could be replaced by a qualitatively different object, such as a wormhole \cite{Damour:2007ap,Cardoso:2016rao,Konoplya:2016hmd,Lemos:2008cv,Bronnikov:2019sbx,Guerrero:2022qkh,Yang:2021cvh,Churilova:2019cyt} or a Schwarzschild star \cite{Konoplya:2019nzp}, with essentially the same basic astrophysical observable quantities. However, such exotic objects not only have not been observed so far, but usually have internal theoretical contradictions such as instability against small perturbations or requirement of various forms of the exotic matter, etc. At the same time, there must be a well-justified mimicker of the Schwazrschild/Kerr black hole -- another black hole which is Schwarzschild-like everywhere, except a small region near the event horizon, where, the deviation from the Einsteinian gravity is considerable.
Such near-horizon deformations can be induced by quantum corrections or possible extra dimensional scenarios. This kind of a black hole would look for current and near future experiments \cite{LIGOScientific:2016aoc,EventHorizonTelescope:2019dse} as essentially Einsteinian. However, being able to deform the geometry near the horizon one can change considerably the temperature of such a black hole. Then we would have a model of the black hole possessing basic classical properties indistinguishable from the Schwarzschild ones, but clearly distinctive Hawking radiation, which is strongly determined by the temperature of the event horizon.

Finding of such a toy model mimicker would be an easy task and, as a matter of fact, there are various black-hole solutions already possessing this property at least in some range of the parameters (see, for example, \cite{Konoplya:2022hll,Konoplya:2023aph}). A more interesting problem is to construct a sufficiently {\it general} black-hole spacetime mimicking classical observable effects of the Einsteinian black hole, but leaving a window for essential non-Einsteinian near-horizon geometry and consequently different Hawking radiation. However, the classical (quasinormal) spectrum of such a black hole is expected to be different from the Schwarzschild one. As was shown in \cite{Konoplya:2022hll} and checked in numerous examples \cite{Konoplya:2022pbc,Konoplya:2023ppx,Konoplya:2022iyn}, when the metric differs noticeably from the Schwarzschild one only in the near-horizon zone, the least damped (fundamental) quasinormal mode is indeed very close to its Schwarzschild value, but the several first overtones may deviate a lot, by hundreds of percent. Such an outburst of overtones was observed earlier when the effective potential is slightly perturbed in the whole space \cite{Jaramillo:2020tuu}. In \cite{Konoplya:2022pbc} it was shown that a tiny deformation solely in the near horizon zone is sufficient for significant change of overtones' frequencies. Moreover, the astrophysically relevant perturbations of the geometry in the far zone do not lead to a similar effect, so that the overtones probe the event-horizon geometry \cite{Konoplya:2022pbc}. Since the overtones are necessary in order to reproduce the earlier period of quasinormal ringing \cite{Giesler:2019uxc}, one can expect their detection with the next generation of gravitational wave antennas \cite{Oshita:2022yry}.

In order to construct the aforementioned general black-hole spacetime we will use the general parametrization of spherically symmetric and asymptotically flat black holes in arbitrary metric theories of gravity \cite{Rezzolla:2014mua}. The parametrization is based on a continued-fraction expansion in terms of a compactified radial coordinate, which leads to superior convergence properties and strict hierarchy of coefficients, allowing us to approximate this or that black-hole metric with a much smaller number of coefficients. Generally, the parametrization contains an infinite number of coefficients, but only a few of them are important for basic observable quantities \cite{Konoplya:2020hyk,Konoplya:2022tvv}, so that the infinite continued fraction can usually be safely truncated at the first few orders of expansion.
We will consider various essential characteristics of the parametrized black hole, quasinormal modes (QNMs), radius of the shadow, frequency at the innermost stable circular orbit (ISCO), binding energy of particle, and require they must be practically indistinguishable from those for the Schwarzschild black hole. In addition, we will require that the post-Newtonian (PN) constraints are fulfilled in the far zone. This will lead to a compact black-hole metric depending on the five parameters, which, under reasonable assumptions can be reduced to the three independent parameters.

The metric has the form:
\begin{equation}
ds^2=-N^2(r)dt^2+\frac{B^2(r)}{N^2(r)}dr^2+r^2 (d\theta^2+\sin^2\theta d\phi^2),\label{metric}
\end{equation}
where
\begin{eqnarray}\nonumber
N^2(r)&=&\left(1-\frac{r_0}{r}\right) \left(1-\frac{r_0 \epsilon }{r}-\frac{r_0^2 \epsilon }{r^2}+\frac{r_0^3}{r^2} \frac{a_1}{r+a_2(r-r_0)}\right),\!\!\!\!\\
\label{metricf}
B^2(r)&=&\left(1+\frac{r_0^2}{r}\frac{b_1 }{r+b_2(r-r_0)}\right)^2.
\end{eqnarray}
In addition, supposing that deformations are localized solely in the near horizon zone (for example, due to quantum corrections or extra dimensional scenarios), we will require that the eikonal quasinormal modes coincide with the Schwarzschild ones. That is a reasonable condition, because the eikonal frequencies are determined by the geometry near the peak of the effective potential, i.e. at a distance from the horizon, where the near-horizon effects must be completely negligible. Then, for sufficiently small values of the deformation parameter $\epsilon$, which quantifies how the black-hole radius differs from its Schwarzschild value, coefficients $a_1$ and $b_1$ are expressed in a compact form in terms of $a_2>-1$ and $b_2>-1$,
\begin{equation}\label{1storder}
\begin{array}{rcl}
a_1&=&-(3+a_2)\epsilon,\\\mbox{}\\
b_1&=&-\dfrac{4(2+a_2)(3+b_2)}{(3+a_2)^2}\epsilon.
\end{array}
\end{equation}
This way, fixing the radius of the black hole $r_{0}$, we obtain our main result here: a general metric describing a Schwarzschild mimicker which depends {\it only on the three parameters} $\epsilon$, $a_2$ and $b_2$. The parameter $\epsilon$ is the measure of deviation of the black-hole radius from its Schwarzschild radius $2 M$, while the coefficients $a_2$ and $b_2$ determine the geometry of the black hole in the proximity of its event horizon.

The paper is organized as follows. In Sec.~\ref{sec:parametrization} we briefly relate the general parametrization for spherically symmetric black holes. Sec.~\ref{sec:particlemotion} is devoted to particle motion around such parametrized black hole. Sec.~\ref{sec:constraints} discusses the constraints on the coefficients of the parametrization. In Sec.~\ref{sec:QNMs} we study quasinormal modes of these black holes.
In the Conclusions we summarize the obtained results and discuss possible directions of future research.

\section{The continued fraction expansion}\label{sec:parametrization}
We consider general form of the line element describing a spherically symmetric black hole (\ref{metric}) with the event horizon radius $r_0$, so that $N^2(r_0)=0$.

Following \cite{Rezzolla:2014mua}, we use the dimensionless variable
$$x \equiv 1-\frac{r_0}{r},$$
so that $x=0$ corresponds to the event horizon, while $x=1$ corresponds to spatial infinity. In addition, we rewrite the metric function $N$ as
$$N^2=x A(x),$$
where $A(x)>0$ for \mbox{$0\leq x\leq1$}.
Using the new parameters $\epsilon$, $a_0$, and $b_0$, the functions $A$ and $B$ can be written as
\begin{eqnarray}\nonumber
A(x)&=&1-\epsilon (1-x)+(a_0-\epsilon)(1-x)^2+{\tilde A}(x)(1-x)^3\,,
\\
B(x)&=&1+b_0(1-x)+{\tilde B}(x)(1-x)^2\,.
\end{eqnarray}
Here the coefficient $\epsilon$ measures the deviation of $r_0$ from $2M$,
$$\epsilon = \frac{2 M-r_0}{r_0}.$$
The coefficients $a_0$ and $b_0$ can be seen as combinations of the PPN parameters:
$$a_0=\frac{(\beta-\gamma)(1+\epsilon)^2}{2}, \qquad b_0=\frac{(\gamma-1)(1+\epsilon)}{2}.$$
Current observational constraints on the PPN parameters imply that the coefficients $a_0$, $b_0$ are of the order \mbox{$a_0 \sim b_0 \sim 10^{-4}$} or smaller.

The functions ${\tilde A}$ and ${\tilde B}$ are introduced through infinite continued fraction in order to describe the metric near the horizon (\ie for $x \simeq 0$),
\begin{equation}\label{ABdef}
{\tilde A}(x)=\frac{a_1}{\displaystyle 1+\frac{\displaystyle a_2x}{\displaystyle 1+\ldots}}, \qquad
{\tilde B}(x)=\frac{b_1}{\displaystyle 1+\frac{\displaystyle b_2x}{\displaystyle1+\ldots}},
\end{equation}
where $a_1, a_2,\ldots$ and $b_1, b_2,\ldots$ are dimensionless constants to be constrained from observations of phenomena near the event horizon. At the horizon only the first two terms of the expansions survive,
$
{\tilde A}(0)={a_1},~
{\tilde B}(0)={b_1},
$
which implies that near the horizon only the lower order terms of the expansions are important.

\section{Particle motion}\label{sec:particlemotion}
We shall use the auxiliary function \cite{Konoplya:2020hyk},
\begin{equation}\label{auxiliary}
P(x)\equiv (1-x)^2xA(x),
\end{equation}
allowing one to identify the {\it radiation zone}, which is conditionally interpreted as a region where classical radiation processes are significant. The latter implies that the near horizon layer and the asymptotic zone (sufficiently far from the innermost stable circular orbit) are excluded from the radiation zone.

Defining the four-momentum of a particle of mass $m$ as
\begin{equation}\label{momentum}
p^{\mu}\equiv m\frac{dx^{\mu}}{ds},
\end{equation}
where $s$ is an invariant affine parameter, and introducing the energy $E \equiv -p_t$ and angular momentum $L\equiv p_{\phi}$, from the normalization condition on the four-momentum we have
\begin{equation}\label{normalization}
p_{\mu}p^{\mu}=-m^2.
\end{equation}
Due to spherical symmetry, without loss of generality we consider the motion in the equatorial plane $\theta=\pi/2$, $d\theta=0$ and obtain the following equation for the radial coordinate
\begin{equation}\label{radialeq}
m^2 g_{rr}\left(\frac{dr}{ds}\right)^2=V_{eff},
\end{equation}
where the effective potential is determined as
\begin{eqnarray}\label{potential}
V_{eff}&=&-(g^{tt}E^2+g^{\phi\phi}L^2+m^2)
\PRDstyle{\\\nonumber&=&}
\JCAPstyle{=}
\frac{E^2}{N^2(r)}-\frac{L^2}{r^2}-m^2
\\\nonumber
&=&(1-x)^2\left(\frac{E^2}{P(x)}-\frac{L^2}{r_0^2}\right)-m^2.
\end{eqnarray}

The circular motion corresponds to the constant radial coordinate, so that $dr=0$ and $d^2r=0$, which is equivalent to
\begin{equation}\label{circulareq}
\begin{array}{rcl}
V_{eff}(x) &=& 0, \\
V_{eff}'(x) &=& 0.
\end{array}
\end{equation}

Then, $E$ and $L$ at the given circular orbit are
\begin{equation}\label{circularEL}
\begin{array}{rcl}
E^2&=&-m^2\dfrac{2P^2(x)}{(1-x)^3P'(x)},\\
L^2&=&m^2r_0^2\dfrac{2P(x)+(1-x)P'(x)}{(1-x)^3P'(x)},\\
\end{array}
\end{equation}
and the corresponding frequency of rotation is defined as
\begin{equation}\label{frequency}
\Omega^2\equiv\left(\frac{d\phi}{dt}\right)^2=\frac{L^2N^4(r)}{E^2r^4}=\frac{2P(x)+(1-x)P'(x)}{2r_0^2}.
\end{equation}

For the massless particle \mbox{($m=0$)} Eqs.~(\ref{circulareq}) have only one solution, corresponding to the orbit, for which $P(x)$ attains its maximum. The shadow radius is defined by the impact parameter $R_s=L/E$ of the photons which leave the proximity of the circular orbit.
Therefore, $R_s$ can be determined by finding the maximum value of $P(x)$,
\begin{equation}\label{shadow}
\frac{r_0^2}{R_{s}^2}=\max P(x)=P(x_m),
\end{equation}
where we took into account that $P'(x_m)=0$.

After substituting \mbox{$ds^2=m^2g_{tt}^2dt^2/E^2$} into Eq.~(\ref{radialeq}), taking the limit \mbox{$m\to0$}, and substituting $x=x_m+\delta x$, one finds the equation for the radial coordinate of such photons,
\begin{equation}\label{LypunovEq}
\left(\frac{d}{dt}\delta x\right)^2=\lambda^2\delta x^2+\Order{\delta x}^3,
\end{equation}
where $\lambda$ is the Lyapunov exponent, which can be calculated, using the following relation:
\begin{equation}\label{Lyapunov}
\lambda^2 = -\frac{P(x_m)P''(x_m)}{2r_0^2B^2(x_m)}.
\end{equation}

Usually, the eikonal regime ($\ell \rightarrow \infty$) of QNMs \cite{eikonal,Schutz:1985km} can be represented through these quantities as follows \cite{Cardoso:2008bp}:
\begin{equation}\label{eikonal}
\omega=\frac{1}{R_{s}}\left(\ell+\frac{1}{2}\right)-\imo\lambda\left(n+\frac{1}{2}\right)+\Order{\frac{1}{\ell}}.
\end{equation}
Though there exist a number of exception from this correspondence between the eikonal QNMs and the characteristics of the null geodesics \cite{Khanna:2016yow,Konoplya:2017wot,Konoplya:2022gjp}.

For the Schwarzschild black hole we have
\begin{equation}
x_m=\frac{1}{3}, \qquad R_{s}=\lambda^{-1}=\frac{3\sqrt{3}}{2}r_0=3\sqrt{3}M.
\end{equation}

For a massive particle among the family of the solutions to (\ref{circulareq}) there is the innermost stable circular orbit (ISCO), corresponding to the minimum of energy,
\begin{equation}\label{EISCO}
E_{ISCO}^2=-m^2\max\dfrac{2P^2(x)}{(1-x)^3P'(x)},
\end{equation}
which is attained at the orbit of $x=x_{ISCO}$.

The invariant characteristics of the ISCO are its frequency $\Omega_{ISCO}$, calculated using (\ref{frequency}) for $x=x_{ISCO}$ and the binding energy, \ie the amount of energy per unit mass released by a particle going over from the distant orbit with $E\simeq m$,
\begin{equation}\label{BEdef}
BE=\frac{E-E_{ISCO}}{m}=1-\frac{E_{ISCO}}{m}.
\end{equation}

In the Schwarzschild limit, these quantities take the following values
\begin{eqnarray}
x_{ISCO}=\frac{2}{3},\qquad BE=1-\frac{2\sqrt{2}}{3},
\PRDstyle{\\\nonumber}
\JCAPstyle{\qquad}
\Omega_{ISCO}=\frac{\sqrt{6}}{18r_0}=\frac{\sqrt{6}}{36M}.
\end{eqnarray}

\section{Constraining the coefficients}\label{sec:constraints}
In \cite{Konoplya:2020hyk} we have shown that only several dominant parameters are important for the essential astrophysically observable quantities, such as fundamental quasinormal modes, parameters of the shadow, frequency at ISCO, binding energy etc. Therefore, we shall limit our consideration by the second order in the continued-fraction expansion (\ref{ABdef}), \ie we take $a_3=b_3=0$. This way the following 8 parameters remain: $r_{0}$, $\epsilon$, $a_0$, $b_0$ $a_1$, $a_2$, $b_1$, $b_2$.

\begin{enumerate}
\item
The value of $\epsilon$ is, strictly speaking, constrained only from below. Indeed it is related to the compactness of the object $M/r_{0}$ and if $\epsilon \lesssim - 0.6$, neutron stars would collapse (see \cite{Konoplya:2022tvv} for details). However, if we require that the parametrized metric reproduces the Schwarzschild radius of shadow with sufficient accuracy (see fig. 1), then $\epsilon$ has to be small. Even in a broader context, a great number of black holes which are quite different from the Schwarzschild spacetime normally have $|\epsilon| \lesssim 1$ \cite{Konoplya:2020hyk}.

\item The coefficients $a_0$ and $b_0$ are very small due to constraints on the 1PN parameters. In this paper we shall neglect these coefficients, \ie we consider
\begin{equation}
a_0=b_0=0.
\end{equation}

\item
Current estimations for the 2PN parameters from the binary black hole gravitational signals \cite{LIGOScientific:2019fpa} suggest that $|{\tilde A}(1)|\lesssim1$ and $|{\tilde B}(1)|\lesssim1$, implying that
\begin{equation}\label{1constraint}
|a_1|\lesssim 1+a_2, \qquad |b_1| \lesssim 1+b_2.
\end{equation}

\item
Requirement that within the second-order approximation the metric has no singular points outside the event horizon leads to the following constraints
\begin{equation}\label{2constraint}
a_2>-1,\qquad b_2>-1.
\end{equation}
\end{enumerate}

Thus, we shall study the five-parametric family of spherically symmetric black holes, defined by the coefficients $\epsilon,a_1,b_1,a_2,b_2$, which obey the constraints (\ref{1constraint}) and (\ref{2constraint}).

We will imply that in the background of the mimicker the basic characteristics of the null circular geodesics, such the Lyapunov exponent and the rotation frequency are the same as that of the Schwarzschild spacetime. Consequently, the shadow cast by the mimicker will be indistinguishable from the Schwarzschild one. Owing to the correspondence between null geodesics and eikonal quasinormal modes this requirement will also provide some features of the quasinormal spectrum \cite{Cardoso:2008bp,Khanna:2016yow,Konoplya:2017wot,Konoplya:2022gjp}. Indeed, for a mimicker we assume that the dominant quasinormal modes of the black hole do not significantly differ from the Schwarzschild ones.

Let us formulate this requirement more rigorously. The general, agnostic approach we use does not allow us to calculate quasinormal modes of gravitational perturbations, because we do not fix a particular theory of gravity. Instead we can find frequencies of test fields, which, although usually they deviate from their Schwarzschild values at the same rate \cite{Volkel:2019muj}, differ from the modes of gravitational perturbations. However, in the eikonal regime of large multipole numbers $\ell$ quasinormal modes of test and gravitational fields normally coincide \cite{Cardoso:2008bp,Khanna:2016yow,Konoplya:2017wot}. Moreover, the eikonal formula (\ref{eikonal}) already for moderate and small $\ell$ provides if not a good approximation, then, at least, the same order of deviation of the fundamental mode from its Schwarzschild limit. Therefore, we expect that, if the values of the shadow radius and Lyapunov exponent coincide with the Schwarzschild ones, the dominant quasinormal modes for the test fields also lay quite closely to the Schwarzschild values (see, \eg \cite{Volkel:2022aca,Volkel:2022khh}), what is actually confirmed in numerous examples. Thus, we require that
\begin{equation}
R_{s}=\lambda^{-1}=3\sqrt{3}M.
\end{equation}
These two conditions allow us to fix two of the free coefficients.

Namely, we shall define the values of $a_1$ and $b_1$ in such a way that the shadow radius and Lyapunov exponent coincide with the Schwarzschild values. Then, from (\ref{shadow}) and (\ref{Lyapunov}) we obtain the set of equations
\begin{eqnarray}
\nonumber
P'(x_m)&=&0,\\
\label{a1b1fix}
P(x_m)(1+\epsilon)^2&=&\frac{4}{27},\\
\nonumber
-\frac{P(x_m)P''(x_m)}{2B^2(x_m)}(1+\epsilon)^2&=&\frac{4}{27},
\end{eqnarray}
which let us find $a_1$ and $b_1$ numerically for given values of $\epsilon$, $a_2$ and $b_2$.

For small values of $\epsilon$ we find:
\begin{eqnarray}\label{xexp}
x_m&=&\frac{1}{3}+\frac{2(19+9a_2)}{27(3+a_2)}\epsilon+\Order{\epsilon}^2,\\
\label{a1exp}
a_1&=&-(3+a_2)\epsilon+\Order{\epsilon}^2,\\
\label{b1exp}
b_1&=&-\frac{4(2+a_2)(3+b_2)}{(3+a_2)^2}\epsilon+\Order{\epsilon}^2.
\end{eqnarray}
The Mathematica\textregistered{} notebook with the derivation of the formula for arbitrary order in $\epsilon$ is available from \cite{Notebook}. Notice that $b_1$ depends linearly on $b_2$, therefore error of the approximation due to expansion in terms of $\epsilon$ is of the same order for any value of $b_2$.

From (\ref{a1exp}) and (\ref{b1exp}) it is easy to see that for the nonzero value of $\epsilon$, the coefficient $a_1$ remains finite when \mbox{$a_2\to-1$}, and $b_1$ is finite as \mbox{$b_2\to-1$}. Therefore, in order to satisfy Eq.~(\ref{1constraint}), one should consider values of $a_2$ and $b_2$ sufficiently larger than $-1$. Apart from this region we have not found violations of the constraint (\ref{1constraint}) for the black-hole configurations defined through (\ref{a1b1fix}).

Another restriction to the allowed values of the coefficients comes from the requirement that $A(x)$ and $B(x)$ must be positive definite for $0\leq x\leq1$. It turns out that for large values of $a_2$ or $b_2$ the conditions (\ref{a1b1fix}) lead to either a black hole with the event horizon at some point $r_h>r_0$ ($x_h>0$), which is beyond our initial assumption and means that the proper parametrization for such a configuration corresponds to a different set of the coefficients, or to a naked singularity, which we do not study in the present paper. Therefore, we can exclude both cases from our consideration by imposing the condition $A(0)>0$ and $B(0)>0$, or, equivalently,
\begin{equation}\label{horizoncond}
1+a_1>2\epsilon, \qquad 1+b_1>0,
\end{equation}
where $a_1$ and $b_1$ depend on $a_2$ and $b_2$.

\begin{figure*}
\resizebox{\linewidth}{!}{\includegraphics*{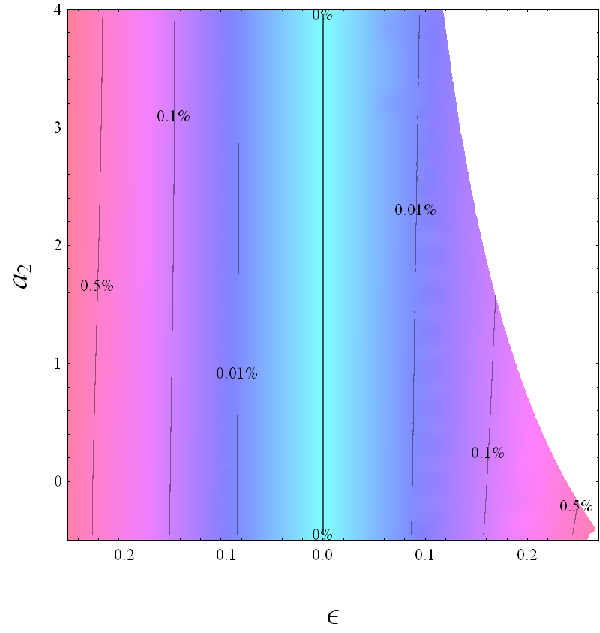}\includegraphics*{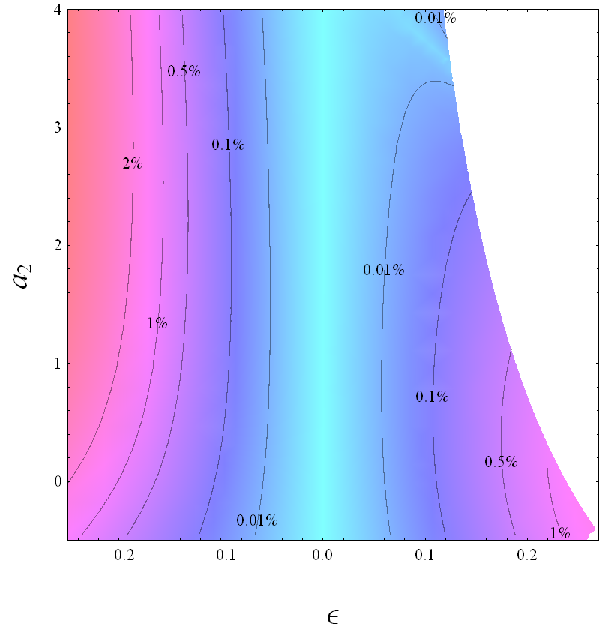}}
\caption{Relative deviation of the shadow radius (left panel) and Lyapunov exponent (right panel) for $b_2=3$ from their Schwarzschild values when $a_1$ and $b_1$ are approximated by the third-order expansion with respect to $\epsilon$. The empty region for large positive values of $\epsilon$ corresponds to the parametric region excluded due to the conditions (\ref{horizoncond}).}\label{fig:3oerror}
\end{figure*}

From Fig.~\ref{fig:3oerror} we can see that the approximation for $a_1$ and $b_1$ in the form of expansion in powers of $\epsilon$ up to the order $\Order{\epsilon}^3$ already provides quite a good mimic of the Schwarzschild black hole in the radiation zone. In order to find significant deviation from the Schwarzschild case relatively large values of $\epsilon$ are necessary: If $|\epsilon| \approx 0.25$, then the deviation of the values of $R_{s}$ is within fractions of a percent, while the deviation of $\lambda$ is a few percent.

\begin{figure*}
\resizebox{\linewidth}{!}{\includegraphics*{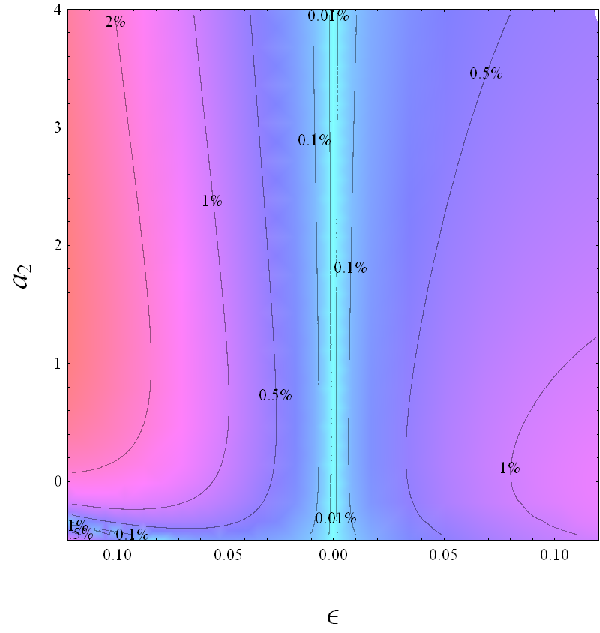}\includegraphics*{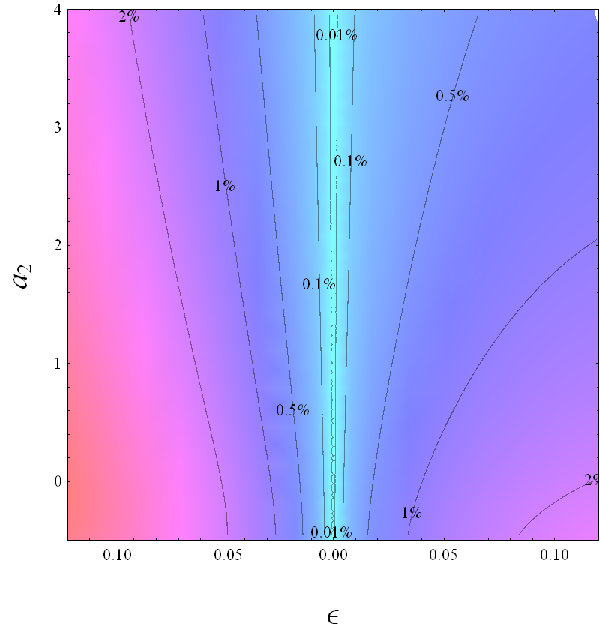}}
\caption{Relative deviation of the ISCO frequency (left panel) and binding energy (right panel) from their Schwarzschild values when $a_1$ is fixed in such a way that the shadow radius and Lyapunov exponent coincide with the Schwarzschild values.}\label{fig:ISCOerror}
\end{figure*}

\begin{figure}
\resizebox{\linewidth}{!}{\includegraphics*{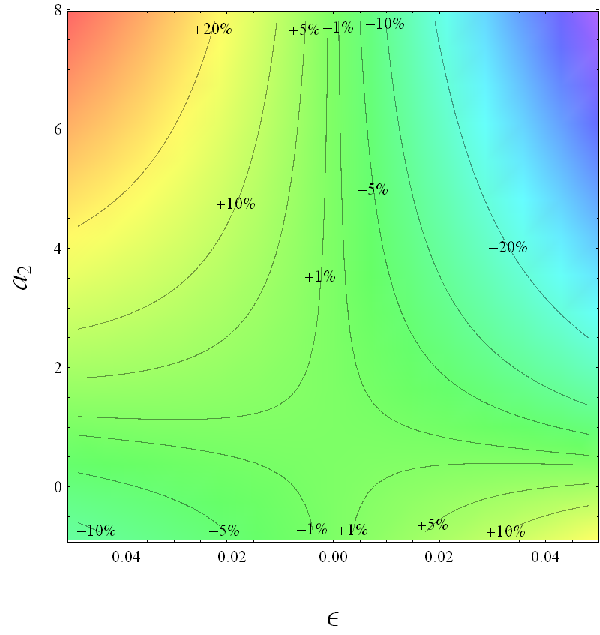}}
\caption{Relative difference between the black-hole Hawking temperature and the Schwarzschild temperature for small values of $\epsilon$ ($b_2=3$).}\label{fig:temperature}
\end{figure}

Since the resulting black hole mimics the Schwarzschild metric both in the radiation zone and in the asymptotic region, it is natural that the dynamics of the massive particles does not deviates from the Schwarzschild limit much as well. From Fig.~\ref{fig:ISCOerror} we see that the deviations of the ISCO frequency and binding energy from the Schwarzschild black hole values are again within several percent for the considered family of the spherically symmetric black holes. Thus, we conclude that the accretion process onto such black holes would give us the same picture of observation data and the only distinction can be observed from the near-horizon phenomena, such as Hawking radiation (for small black holes) or the several lowest overtones in the ringdown (for astrophysical black holes). The reason for the first is that, within our approach, the Hawking temperature obeys the relation
\begin{eqnarray}\label{Hawking}
8\pi MT_H&=&\frac{(1+\epsilon)A(0)}{B(0)}=\frac{(1+\epsilon)(1 + a_1 - 2 \epsilon)}{1 + b_1}\\\nonumber
&=&1+\left(4\frac{(2+a_2)(3+b_2)}{(3+a_2)^2}-4-a_2\right)\epsilon+\Order{\epsilon}^2,
\end{eqnarray}
and although its deviation is proportional to the small parameter $\epsilon$, it can deviate significantly from the Schwarzschild temperature (see Fig.~\ref{fig:temperature}) due to dependence on $a_i$ and $b_i$.

In order to check the validity of the third-order approximation by the series expansion in terms of $\epsilon$ we compare the relative deviation of the frequency at ISCO and binding energy with the deviation of these parameters without using the approximation (Fig.~\ref{fig:ISCOerror}). We conclude that the third-order approximation provides a sufficiently accurate analytic form for the general spacetime mimicking the Schwarzschild black hole. For small values of $|\epsilon|\lesssim0.01$ the first-order approximation (\ref{1storder}) is sufficient for practical purposes.

If we suppose that the near horizon corrections do not decay quickly and still slightly modify the geometry in the radiation zone, which is the usual case for various alternative theories of gravity, then the eikonal quasinormal frequencies and, consequently, radius of the shadow slightly differ from their Schwarzschild values. This slight difference may be lower than the observational precision, so that the spacetime will keep being a Schwarzschild mimicker. In that case the initial form of the metric (\ref{metricf}) with five parameters could be a good approximation
\begin{equation}\label{fivepars}
\begin{array}{rcl}
a_1&=&-(3+a_2)\epsilon + \delta_1,\\\mbox{}\\
b_1&=&-\dfrac{4(2+a_2)(3+b_2)}{(3+a_2)^2}\epsilon + \delta_2.
\end{array}
\end{equation}
Indeed, black-hole metrics in various theories of gravity could be obtained from such a general parametrized form in the regime of moderate deviations from the Schwarzschild spacetime (see \cite{Konoplya:2020hyk,Konoplya:2022tvv} for details). Supposing that the spacetime (\ref{fivepars}) still mimics Schwarzschild characteristics with reasonable accuracy, $\delta_1$ and $\delta_2$ must be sufficiently small.

A good example of such a spacetime is the black hole in the Einstein-scalar-Gauss-Bonnet theory \cite{Konoplya:2019fpy}. For the quadratic scalar coupling,
\begin{eqnarray}\nonumber
\epsilon &=& \frac{\frac{p}{43}-\frac{p^2}{201}}{1-\frac{105 p}{577}}\,,\\
\nonumber
a_1 &=& \frac{\frac{63 p}{332}-\frac{23 p^2}{143}}{-\frac{83 p^2}{401}+p-\frac{223}{259}}\,,\\
a_2 &=& \frac{-\frac{234 p^2}{307}+\frac{152 p}{397}+1}{\frac{73}{221}-\frac{73 p}{228}}\,,\\
\nonumber
b_1&=&\frac{\frac{91 p}{396}-\frac{85 p^2}{438}}{-\frac{24 p^2}{131}+p-\frac{632}{707}}\,,\\
\nonumber
b_2&=&\frac{-\frac{173 p^3}{432}+\frac{47 p^2}{225}-\frac{106 p}{203}+1}{\frac{20}{221}-\frac{25 p}{283}}\,,
\end{eqnarray}
where the parameter $p\in[0,1]$ is proportional to the square of the coupling constant $\alpha$,
$$p=384\frac{\alpha^2\phi(r_0)^2}{r_0^4}.$$
Here $\phi(r_{0})$ is the value of the scalar field at the event horizon (see \cite{Konoplya:2019fpy} for more details). It follows that, for small deviation parameter, $p\ll1$, we have
\begin{equation}
\delta_1=-0.080p+\Order{p^2},\quad\delta_2=-0.076p+\Order{p^2}.
\end{equation}

Similarly, for the cubic coupling, we have
$$p=864\frac{\alpha^2\phi(r_0)^4}{r_0^4},$$
\begin{eqnarray}\nonumber
\epsilon &=& \frac{-\frac{p^3}{186}+\frac{7 p^2}{234}+\frac{3 p}{409}}{p+\frac{34}{271}}\,,\\
\nonumber
a_1 &=& \frac{\frac{11 p}{84}-\frac{79 p^2}{638}}{-\frac{124 p^2}{405}+p-\frac{233}{328}}\,,\\
a_2 &=& \frac{-\frac{179 p^3}{182}+p^2+\frac{97 p}{253}}{-\frac{41 p^2}{212}+\frac{41 p}{209}+\frac{1}{360}}\,,\\
\nonumber
b_1 &=& \frac{\frac{33 p}{145}-\frac{58 p^2}{301}}{-\frac{77 p^2}{414}+p-\frac{217}{244}}\,,\\
\nonumber
b_2 &=& \frac{p^3-\frac{119 p^2}{205}-\frac{186 p}{227}-\frac{1}{123}}{\frac{19 p^2}{181}-\frac{86 p}{811}-\frac{1}{512}} \,.
\end{eqnarray}
Then we find that
\begin{equation}
\delta_1=-0.009p+\Order{p^2},\quad\delta_2=0.116p+\Order{p^2}.
\end{equation}

Since $\delta_1$ and $\delta_2$ are small corrections to the coefficients $a_1$ and $b_1$, respectively, the radiation processes in the vicinity of the Einstein-scalar-Gauss-Bonnet black hole are similar to the Schwarzschild one for small and even moderate values of the coupling parameter $p$.

\section{Quasinormal modes}\label{sec:QNMs}
Here, for illustration, we will consider quasinormal modes of a test electromagnetic field.
The general covariant equations for the four-potential $A_\mu$ has the form:
\begin{subequations}\label{coveqs}
\begin{eqnarray}\label{KGg}
\frac{1}{\sqrt{-g}}\partial_{\mu} \left(F_{\rho\sigma}g^{\rho \nu}g^{\sigma \mu}\sqrt{-g}\right)&=&0\,,
\end{eqnarray}
\end{subequations}
where $F_{\mu\nu}=\partial_\mu A_\nu-\partial_\nu A_\mu$ is the electromagnetic tensor.
After separation of the variables equations Eq. (\ref{coveqs}) takes the Schrödinger wave-like form (see, for instance, \cite{Konoplya:2011qq,Kokkotas:1999bd} and references therein)
\begin{equation}\label{wave-equation}
\dfrac{d^2 \Psi}{dr_*^2}+(\omega^2-V(r))\Psi=0,
\end{equation}
where the ``tortoise coordinate'' $r_*$ is defined by the following relation
\begin{equation}
dr_*\equiv\frac{B(r)dr}{N^2(r)}.
\end{equation}

Quasinormal modes are proper oscillation frequencies of a black hole gravitational field or test fields in its vicinity, corresponding to the following boundary conditions: purely ingoing waves at the event horizon ($r_* \to -\infty$) and purely outgoing wave at infinity ($r_* \to +\infty$). Thus, in terms of the tortoise coordinate no waves from either plus or minus infinity are coming. In order to find quasinormal modes accurately we use the Leaver method \cite{Leaver:1985ax}, which is based on the convergent procedure using the Frobenius series expansion. The wave-like equation~(\ref{wave-equation}) has regular singularities at the event horizon $r=r_0$ and the irregular singularity at spatial infinity $r=\infty$. We introduce the new function
\begin{equation}\label{reg}
\Psi(r)=e^{\imo\omega r}r^{\lambda}\left(1-\frac{r_0}{r}\right)^{-\imo\omega/4\pi T_H}y(r),
\end{equation}
where $\lambda$ is defined in such a way that $y(r)$ is regular at $r=\infty$ once $\Psi(r)$ corresponds to the purely outgoing wave at spatial infinity. We notice that $\Psi(r)$ behaves as the purely ingoing wave at the event horizon if $y(r)$ is regular at $r=r_+$. Therefore, we represent $y(r)$ in terms of the following Frobenius series:
\begin{equation}\label{Frobenius}
y(r)=\sum_{k=0}^{\infty}a_k\left(1-\frac{r_0}{r}\right)^k,
\end{equation}
and find that the coefficients $a_k$ satisfy the n-term recurrence relation, which can be reduced to the three-term recurrence relation via Gaussian elimination (see, for example, \cite{Konoplya:2011qq} for details). Then, using the coefficients in the recurrence relation, we find an infinite continued fraction equation with respect to $\omega$, which is satisfied when the series (\ref{Frobenius}) converges at $r=\infty$, or, in other words, if $\Psi(r)$ satisfies the quasinormal boundary conditions. In order to obtain the final equation we employ a sequence of positive real midpoints as described in \cite{Rostworowski:2006bp} and use the Nollert improvement~\cite{Nollert:1993zz}, which was generalized in~\cite{Zhidenko:2006rs} for the recurrence relation of arbitrary number of terms.

\PRDstyle{\begin{table*}}
\JCAPstyle{\begin{sidewaystable}}
\begin{tabular}{|l|c|c|c|c|c|}
\hline
  $n$ & $\epsilon=-0.04$ & $\epsilon=-0.02$ & $\epsilon=0$ & $\epsilon=0.02$ & $\epsilon=0.04$ \\
\hline
  $0$ & $0.248974-0.095166\imo$ & $0.248783-0.093811\imo$ & $0.248263-0.092488\imo$ & $0.247452-0.091469\imo$ & $0.246637-0.090964\imo$ \\
  $1$ & $0.221900-0.308324\imo$ & $0.219740-0.300992\imo$ & $0.214515-0.293668\imo$ & $0.205868-0.289978\imo$ & $0.201128-0.292654\imo$ \\
  $2$ & $0.203927-0.560494\imo$ & $0.195866-0.543180\imo$ & $0.174774-0.525188\imo$ & $0.144761-0.547283\imo$ & $0.146284-0.540832\imo$ \\
  $3$ & $0.212144-0.827408\imo$ & $0.195728-0.800256\imo$ & $0.146177-0.771909\imo$ & $0.141127-0.810528\imo$ & $0.124068-0.811314\imo$ \\
  $4$ & $0.238569-1.096145\imo$ & $0.213937-1.059562\imo$ & $0.126554-1.022550\imo$ & $0.119833-1.060867\imo$ & $0.118493-1.089125\imo$ \\
  $5$ & $0.275432-1.364445\imo$ & $0.243659-1.318457\imo$ & $0.112253-1.273926\imo$ & $0.123642-1.361626\imo$ & $0.112090-1.362852\imo$ \\
  $6$ & $0.318360-1.631888\imo$ & $0.280402-1.576653\imo$ & $0.101215-1.525266\imo$ & $0.149846-1.608058\imo$ & $0.116237-1.617093\imo$ \\
  $7$ & $0.365168-1.898397\imo$ & $0.321534-1.834164\imo$ & $0.092324-1.776399\imo$ & $0.155416-1.851481\imo$ & $0.134020-1.884589\imo$ \\
  $8$ & $0.414700-2.164090\imo$ & $0.365581-2.091043\imo$ & $0.084935-2.027306\imo$ & $0.146022-2.121117\imo$ & $0.141995-2.152590\imo$ \\
  $9$ & $0.466244-2.429134\imo$ & $0.411693-2.347367\imo$ & $0.078650-2.278009\imo$ & $0.187734-2.390474\imo$ & $0.153225-2.406932\imo$ \\
\hline
\end{tabular}
\caption{Dominant modes for electromagnetic perturbations ($\ell=1$) of the Schwarzschild black hole mimickers $M=1$, $a_2=6$, $b_2=3$ approximated by the third-order expansion with respect to $\epsilon$ ($\epsilon=0$ corresponds to the Schwarzschild black hole).}\label{tabl:QNMs}
\PRDstyle{\end{table*}}
\JCAPstyle{\end{sidewaystable}}

Here we do not have the purpose to study quasinormal modes exhaustively, but rather to show that while the fundamental mode is very close to its Schwarzschild limit within the determined range of parameters, overtones can change at a much higher rate, because of their sensitivity to even small changes of the near horizon geometry \cite{Konoplya:2022pbc}. From the table~\ref{tabl:QNMs} we can see an example of such behavior: variation of $\epsilon$ at a fixed $a_2$ and $b_2$ leads, at maximum, to the two percent deviation of the fundamental mode from the Schwarzschild value ($\epsilon=0$), while several first overtones can deviate by hundreds of percent.

\section{Conclusions}
Here, using reasonable assumptions, such as asymptotic flatness, post-Newtonian, and other straightforward constraints, we have derived a general and simple form of the metric which can mimic the Schwarzschild black hole. When fixing the eikonal quasinormal modes (and consequently the shadow radius) to their Schwarzschild values, the metric depends on the three parameters only appearing as an appropriate truncation of the general parametrization of spherically symmetric and asymptotically flat black holes. In principle, our approach is agnostic and covers arbitrary metric theories of gravity, which reduce to General Relativity in the weak-field regime.

Our approach could be further developed to axially rotating spacetimes in order to construct a general mimicker of the Kerr solution. For this one could use the general axially symmetric parametrization of the black-hole spacetime \cite{Konoplya:2016jvv,Younsi:2016azx}.

It is also tempting to study further properties of such a general mimicker. First of all detailed calculation of quasinormal modes for various spin of field could be done using either Frobenius or other accurate method. Then, the intensity of Hawking radiation both in the semi-classical \cite{Hawking:1975vcx,Page:1976df} and beyond \cite{Parikh:1999mf} regimes could be estimated and more advanced thermodynamic properties studied.

\acknowledgments
A.~Z. was partially supported by Conselho Nacional de Desenvolvimento Científico e Tecnológico (CNPq) and the EU project for development of R\&D capacities of the Silesian University in Opava CZ.02.2.69/0.0/0.0/18\_054/0014696.

\end{document}